# Extended Fokker Planck model: properties and solutions

Sergey Kamenshchikov

*Moscow State University of M.V. Lomonosov, Physical department,
Russia, Moscow, Leninskie Gory, 1/2, 119991. E-mail: kamphys@gmail.com*

**Abstract**

**In the current paper Fokker Planck model of random walks has been extended to non conservative cases characterized by explicit dependence of diffusion and energy on time. A given generalization allows describing of such non equilibrium processes as Levy flights in a classical differential form without use of fractal PDE. Besides it takes into account mixing properties that are obligatory for a certain class of chaotic systems, such as Kolmogorov *K* – system. It was shown that an abnormal transport is a consequence of the equilibrium distortion and not stationary diffusion. The particular case of fixed boundaries was considered. According to the received solutions it was shown that a system structure can resist a weak disturbance in the vicinity of the discrete regimes, defined by a system scale and its nonlinear properties. These regimes correspond to the exponential increase of quasi regular structure fluctuations. Only fast disruption of regime is possible for other states of the system. It leads to an immediate transition to the chaos.**

## 1. Extended Fokker Planck equation

The random walks have been described by several models of mathematical, physical and empirical approaches. Among them the partial differential equation of Fokker-Planck plays a unique role - it allows receiving a wide class of dynamic solutions directly. This model has been modified and finalized by Andrey Kolmogorov [1] in 1938. The basis used for its formulation is Chapman - Kolmogorov equation [2]. If we consider a one dimensional case then a transitional probability for random walks $x_1, t_1 \to x_3, t_3$ formally satisfies the following markovian relation:

$$W(x_3,t_3|x_1,t_1) = \int dx_2 W(x_3,t_3|x_2,t_2) \cdot W(x_2,t_2|x_1,t_1) \qquad (20)$$

In the above formula $x(t)$ is a system coordinate, while $W(x,t|x_0,t_0)$ is a probability density of $x(t)$ location subject to the initial coordinate is $x_0(t_0)$. Fokker – Planck – Kolmogorov (*FPK*) equation has been received based on following assumptions:

**A1**. $W(x',t'|x,t) = W(x',x,t'-t) = W(x',x,t_\Delta)$. It means that a transitional probability doesn't depend on the initial time point. This demand implies that condition of $\tau \succ\succ \tau_C$ is satisfied. $\tau_C$ is an effective width of auto correlation function for x(t). It is also called the auto correlation release time or a system memory loss time. Generally a release of the correlation is caused by the instability of phase trajectories. The memory loss time $\tau_C$ is directly connected with instability characteristic –dynamic entropy $h = \dfrac{\partial S}{\partial t}$. In general case we have the following expression: $\tau_C \sim \dfrac{1}{h}$ [3]. Therefore the existence of $\tau_C$ is provided by the positivity property of $h$ which is a necessary condition of a transition;

**A2.** $P(x',t) = W(x',x,t)$. A final probability doesn't depend on the initial coordinate. This restriction implies $\tau \succ\succ \tau_c$ as well;

**A3** The initial distribution density is defined by Dirac delta function: $W(x) = \delta(0)$ - we assume that the initial coordinate can be defined quite accurately in relation to the general size of a considered system;

**A4.** According to A3 assumption an approximation of the second order may be received for the transitional probability:

$$W(x, x_0, _\Delta t) = \delta(x - x_0) + a(x, _\Delta t) \cdot \delta'(x - x_0) + \frac{1}{2} \cdot b(x, _\Delta t) \cdot \delta''(x - x_0) \quad (21)$$

Factors $a(x, _\Delta t)$ and $b(x, _\Delta t)$ are defined by relations (22) and (23):

$$a(x, _\Delta t) = \int (x - x_0) \cdot W(x, x_0, _\Delta t) dx_0 = \langle\langle _\Delta x \rangle\rangle \quad (22\ a)$$

$$b(x, _\Delta t) = \int (x - x_0)^2 \cdot W(x, x_0, _\Delta t) dx_0 = \langle\langle _\Delta x^2 \rangle\rangle \quad (23\ a)$$

These assumptions form a basis of FPK. Let's introduce a first transport factor $A(x)$ and a second transport factor $B(x)$ in the following way:

$$A(x) = \lim_{_\Delta t \to 0} \left( \frac{\langle\langle _\Delta x \rangle\rangle}{_\Delta t} \right) = \frac{\langle\langle _\Delta x \rangle\rangle}{_\Delta t_{min}} \quad (22\ b)$$

$$B(x) = \lim_{_\Delta t \to 0} \left( \frac{\langle\langle _\Delta x^2 \rangle\rangle}{_\Delta t} \right) = \frac{\langle\langle _\Delta x^2 \rangle\rangle}{_\Delta t_{min}} \quad (23\ b)$$

Then substitution of expansion (21) into Chapman - Kolmogorov equation [2] gives us the following form for $P(x',t) = W(x',x,t)$:

$$\frac{\partial P(x,t)}{\partial t} = \frac{1}{2} \cdot \frac{\partial}{\partial x}\left( B(x) \cdot \frac{\partial P(x,t)}{\partial x} \right) \quad A(x) = \frac{1}{2} \cdot \frac{\partial B(x)}{\partial x} \quad (24)$$

This model uses a second order approximation for the transitional probability (21), as it was stated above in the mathematical form. Here we used a relation between the transport factors.

Let us define a specific energy of a complex system:

$$\varepsilon(x,t) = \lim_{_\Delta t \to 0} \left( \frac{_\Delta x(x,t)}{_\Delta t} \right)^2 \approx \left( \frac{_\Delta x(x,t)}{_\Delta t_{min}} \right)^2 \quad (25)$$

Here $x(t)$ is a characteristic vector in a multidimensional generalized phase space. A second transport factor can be expressed then through this quantity in the following way:

$$B(x,t) \approx \frac{\langle\langle _\Delta x^2 \rangle\rangle}{_\Delta t_{min}} = _\Delta t_{min} \cdot \int \varepsilon(x,t) \cdot W(x, x_0, _\Delta t) dx_0 = _\Delta t_{min} \cdot \langle\langle \varepsilon(x,t) \rangle\rangle \quad (26)$$

For a conservative case an averaged specific energy is presented as: $\langle\langle \varepsilon(x,t) \rangle\rangle = \langle\langle \varepsilon(x) \rangle\rangle$. According to relation (26) this means that transport factor doesn't depend on time explicitly and standard FPK is valid. However this expression is not satisfied if $x(t) \leftrightarrow \varepsilon(t)$ bijection is absent. According to (26) averaged specific energy conservancy may be violated in this case. In general case we should consider a non conservative system and corresponding transport equation. A mutual correspondence is distorted because a phase trajectory mixing occurs - several system states are possible for the same phase space location. A second transport factor $B(x)$ in fact expresses diffusion in phase space, as it was remarked by G.M.Zaslavsky [3]. For the considered type of motion it has explicit time dependence: $B(x,t)$. A non stationary diffusion of trajectories is a common property of many natural transport processes like Levy flights when a clustered phase space is defined [3].

Taking into account the non stationary properties we may modify the assumption A4 in the following way:

$$W(x,x_0,_\Delta t) = \delta(x-x_0) + a(x,_\Delta t,t)\cdot \delta'(x-x_0) + \frac{1}{2}\cdot B(x,_\Delta t,t)\cdot \delta''(x-x_0) \qquad (27)$$

Derivative of a given probability $P(x',t)$ is represented according to the Chapman Kolmogorov equation in the usual form:

$$\frac{\partial P(x,t)}{\partial t} = \lim_{\Delta t \to 0}\left[\frac{1}{\Delta t}\cdot\left(\int dx\cdot P(x_0,t)\cdot(W(x,x_0,_\Delta t) - \delta(x-x_0))\right)\right] \qquad (28)$$

Substitution of (27) into (28) gives non conservative Extended FPK (EFPK):

$$\frac{\partial P(x,t)}{\partial t} = \frac{1}{2}\cdot\frac{\partial}{\partial x}\left(B(x,t)\cdot\frac{\partial P(x,t)}{\partial x}\right) \qquad (29)$$

In multidimensional case we should replace $x$ with vector $\vec{x}$ and differential operator $\frac{\partial}{\partial x}$ with $\sum_i \frac{\partial}{\partial x_i}$. A generalized form then has to be written as it follows below:

$$\frac{\partial P(\vec{x},t)}{\partial t} = \frac{1}{2}\cdot\sum_{i,j}\frac{\partial}{\partial x_i}\left(B_{ij}(\vec{x},t)\cdot\frac{\partial P(\vec{x},t)}{\partial x_j}\right) \qquad (30)$$

Here the second transport factor $B_{ij}(\vec{x}',t)$ is a tensor.

We have extended the approach of Kolmogorov to the arbitrary class of Markov systems. Now this model takes into account mixing properties which are obligatory for many well known chaotic systems, for example $K$ – system of Kolmogorov [4].

## 2. Abnormal transport and Levy flights

Let's consider asymptotic cases of EFPK transport. According to (26) enlarging of time resolution leads to the following relation:

$$B(x,t)\cdot t = \langle\langle_\Delta x^2\rangle\rangle \qquad (t-t_0) \succ\succ _\Delta t_{min} \qquad (31)$$

We may introduce a generalized deviation $\sigma_x = \sqrt{\langle\langle_\Delta x^2\rangle\rangle}$ and the second diffusion $D(x,t) = \sqrt{B(x,t)}$. Then (31) can be modified to the habitual form of a Brown random walk with a non stationary diffusion:

$$\sigma_x = D(x,t)\cdot\sqrt{t} \qquad (32)$$

This type of motion includes a classical case of a Brown particle shift when $D(x,t) = f(x)$ is stationary quantity. However in the other cases a motion may have properties of a random walk and memory properties in the different parts of its trajectory. Indeed if the diffusion change can be neglected the motion stays totally random – changes of coordinates are mutually independent.

However a diffusion instability, for example in case of a power injection (resonances), may lead to Levy flight, when deterministic properties are defined by the first factor of relation (32). In a general case the given expression is called a fractal noise [5] with non zero correlation. Let's approximate diffusion law as a power function: $D(x,t) = k \cdot t^{\frac{\alpha}{2}}$. Then its substitution in (32) and (31) gives us the following expression:

$$\langle\langle_\Delta x^2\rangle\rangle = k^2 \cdot t^\beta \quad \beta = \alpha + 1 \quad k = const \tag{33}$$

Here in general case $\beta$ is a fractional factor. $\alpha \succ 0$ ($\alpha \prec 0$) corresponds to an attractor and a repeller model in terms of a phase area description.

It was mentioned above that initially the properties of motion are defined by an averaged specific energy: $B(x,t) \sim \langle\langle \varepsilon(x,t)\rangle\rangle$. In case of an ergodic system with strong mixing a secondary averaging corresponds simply to mean energy:

$$\langle\langle \varepsilon(x,t)\rangle\rangle = \int \varepsilon(x,t) \cdot W(x,_\Delta t)dx = \overline{\varepsilon(x,t)} \tag{34}$$

Transitional probability here just corresponds to a common probability: $W(x, x_{0,\Delta}t) \approx P(x,t)$. This particular case is realized when a strong mixing of trajectories leads to a fast memory loss. A strange attractor is a typical case. As for ergodicity, it naturally follows from *FPK* model itself.

According to A1 assumption $_\Delta t \succ \tau_C$. Then covariance function $\lim_{t \to 0} K_x(\tau) = 0$ for $t \succ\succ _\Delta t$ and Slutsky criterion of an ergodicity is satisfied automatically:

$$\lim_{T \to \infty}\left[\frac{1}{T}\int_0^T K_x(\tau)\left(1 - \frac{\tau}{T}\right)d\tau\right] = 0 \tag{35}$$

## 3. Phase spectrum of limited system

The definition of the physical sense for $B(x,t)$ factor allows moving on to particular solutions of one dimensional *EFPK*. The simplest case corresponds to a conservative problem of the fixed boundary:

$$\frac{\partial P(x,t)}{\partial t} = \frac{B}{2} \cdot \frac{\partial^2 P(x,t)}{\partial x^2} \tag{36}$$

$$P(0,t) = P(L,t) = 0 \quad x \in [0, L_x]$$

$$P(x,0) = P_0(x)$$

Equation (36) mathematically corresponds to a uniform linear diffusion PDE. Here $B = const$ according to the condition of an energy conservancy – we consider the system's equilibrium phase state:

$$B(x,t) \approx _\Delta t_{min} \cdot \varepsilon \cdot \int_0^{L_x} W(x, x_{0,\Delta}t)dx_0 = \varepsilon \cdot _\Delta t_{min} = const \tag{37}$$

It may be seen that a transport factor finally depends on the accessible time resolution – that's a relativity property of random walks characteristics. The integration into relation (37) is simplified on the basis of the probability normalization condition.

A solution of (36) may be searched in a form of the Fourier expansion:

$$P(x,t) = \sum_{j=1}^{N} c_j(t) \cdot \sin\left[\left(\frac{\pi \cdot j}{L_x}\right) \cdot x\right] \qquad (38)$$

$$c_j(t) = \frac{2}{L_x} \cdot \int_0^{L_x} P(\zeta,t) \cdot \sin\left[\left(\frac{\pi \cdot j}{L_x}\right) \cdot \xi\right] d\xi$$

A substitution of (38) into (36) gives the following superposition of modes:

$$\sum_{j=1}^{N} \sin\left[\left(\frac{\pi \cdot j}{L_x}\right) \cdot x\right] \cdot \left(\frac{\partial c_j(t)}{\partial t} + \frac{B \cdot c_j(t)}{2} \cdot \left(\frac{\pi \cdot j}{L_x}\right)^2\right) = 0 \qquad (39)$$

For an arbitrary $L_x$ a second factor has obligatorily zero value:

$$\frac{\partial c_j(t)}{\partial t} = -\frac{B \cdot c_j(t)}{2} \cdot \left(\frac{\pi \cdot j}{L}\right)^2 \qquad (40)$$

Possible values of a transport factor $B$ may be expressed then in the following way:

$$B_j = -\frac{2}{c_j(t)} \cdot \frac{dc_j(t)}{dt} \cdot \left(\frac{L}{\pi \cdot j}\right)^2 \qquad (41)$$

Let's consider instability of the particles density: $c_j(t) = c_j(0) \cdot \exp(\phi_j \cdot t)$. It leads to a clustering of particles and finally to the non stationary regime when $B = B(t)$. Here $c_j^0 = c_j(0)$ corresponds to an initial density probability, defined by third relation in group (36). Factors $\phi_j$ characterize modes instability increments and depend on nonlinear properties of the complex system media. They are assumed to be constant in given approximation for given complex system. Here we consider the states in the vicinity of a conservative condition - they may be realized during a *phase-phase* transition. Normalization condition for $P(x,t)$ can be represented in the following way:

$$\sum_{j=1}^{N} c_j^0 \cdot \exp(\phi_j \cdot t) \int_0^{L_x} \sin\left[\left(\frac{\pi \cdot j}{L_x}\right) \cdot x\right] dx = 1 \qquad (42)$$

We can remark that satisfaction of this condition is possible only if both relations $\phi_j \succ 0$ and $\phi_j \prec 0$ are valid for the different modes of spectrum. It means that increasing and decreasing quasi regular fluctuations of particles density occur in the media of a complex system.

According to the relation (41) we have discrete spectrum of phases that can be realized for this considered case:

$$B_j = 2 \cdot \phi_j \cdot \left(\frac{L}{\pi \cdot j}\right)^2 \quad \varepsilon_j = \frac{2 \cdot \phi_j}{\Delta t_{min}} \cdot \left(\frac{L}{\pi \cdot j}\right)^2 \quad j = \overline{1,N} \qquad (43)$$

It corresponds to quantization of specific energy, as it follows from (37) and (43) - possible energy of limited system corresponds to the discrete values (43). However, it is important to notice that energy absorption may switch on additional nonlinear rates of freedom, when weak instability of particles density must be described by a more complex model. That's why a distance between discrete levels does not define an absorption property of system in the given case. The basic conclusion which can be made in this chapter is that any limited *FPK* system can be disturbed weakly only in the vicinity of a discrete set of levels which are defined by system's scale and its nonlinear properties. For other energetic states we may receive only fast disruption of a dynamical regime with the following transition to chaos. According to [6] there are two general regimes of stability loss:

a) A soft excitation of quasi regular motion with the following bifurcations and final transition to a strange attractor;

b) A strong breakdown of the stable system state with fast transition to chaos;

We have shown that a limited system structure can resist weak disturbance only in the vicinity of certain regimes, defined by relations (43). These regimes correspond to exponential increase of quasi regular structure fluctuations.

**Conclusions**

1. We have extended an approach of Fokker-Planck-Kolmogorov to the arbitrary class of Markov systems. Now this model takes into account mixing properties which are obligatory for many well known chaotic systems, for example $K$ – system of Kolmogorov;

2. It was shown that a fractal law of complex walks can be derived on the basis of Extended Fokker Planck model. It explains a combination of the determinative fast shifts and random walks though non stationary specific energy of a system;

3. It turned out that any limited equilibrium $FPK$ system can resist a weak disturbance only in the vicinity of the discrete levels. These regimes correspond to the exponential increase of the quasi regular structure fluctuations.